\documentclass[12pt,preprint]{aastex}
\begin{document}

\title{Witnessing a Large-scale Slipping Magnetic Reconnection along a Dimming Channel during a Solar Flare}

\author{Ju Jing\altaffilmark{1}, Rui Liu\altaffilmark{2}, Mark C. M. Cheung\altaffilmark{3}, Jeongwoo Lee\altaffilmark{1, 4}, Yan Xu\altaffilmark{1}, Chang Liu\altaffilmark{1}, Chunming Zhu\altaffilmark{5}, and Haimin Wang\altaffilmark{1}}

\altaffiltext{1}{Center for Solar-Terrestrial Research, New Jersey Institute of Technology, Newark, NJ 07102-1982, USA; ju.jing@njit.edu}

\altaffiltext{2}{CAS Key Laboratory of Geospace Environment, Department of Geophysics and Planetary Sciences, University of Science and Technology of China, Hefei 230026, China}

\altaffiltext{3}{Lockheed Martin Solar and Astrophysics Laboratory, Palo Alto, CA 94304, USA}

\altaffiltext{4}{Department of Physics and Astronomy, Seoul National University, Seoul 08826, Korea}

\altaffiltext{5}{Department of Physics, Montana State University, Bozeman, MT 59717, USA}

\begin{abstract}
We report the intriguing large-scale dynamic phenomena associated with the M6.5 flare~(SOL2015-06-22T18:23) in NOAA active region 12371, observed by RHESSI, Fermi, and the Atmospheric Image Assembly (AIA) and Magnetic Imager (HMI) on the Solar Dynamic Observatory (SDO). The most interesting feature of this event is a third ribbon (R3) arising in the decay phase, propagating along a dimming channel (seen in EUV passbands) towards a neighboring sunspot. The propagation of R3 occurs in the presence of hard X-ray footpoint emission, and is broadly visible at temperatures from
0.6~MK to over 10~MK through the Differential Emission Measure (DEM) analysis. The coronal loops then undergo an apparent slipping motion following the same path of R3, after a $\sim$ 80~min delay. To understand the underlying physics, we investigate the magnetic configuration and the thermal structure of the flaring region. Our results are in favor of a slipping-type reconnection followed by the thermodynamic evolution of coronal loops. In comparison with those previously reported slipping reconnection events, this one proceeds across a particularly long distance ($\sim$60 Mm) over a long period of time ($\sim$50 min), and shows two clearly distinguished phases: the propagation of the footpoint brightening driven by nonthermal particle injection and the apparent slippage of loops governed by plasma heating and subsequent cooling.

\end{abstract}


\section{INTRODUCTION}
Two flare ribbons separating away from the magnetic polarity inversion line (PIL) and the apparent growth of coronal loops have long been observed, and regarded as compelling evidence for the standard magnetic reconnection model (see \citealt{Shibata2011} for a review). This model, however, deals only with a two-dimensional (2D) magnetic configuration with a translational symmetry along the reconnecting X-line. In three-dimensional (3D) solar magnetic field, magnetic reconnection takes place in a more complex configuration and thus the resulting flares may exhibit a variety of forms which cannot be accommodated by the standard reconnection model (\citealt{Aulanier2012, Janvier2017} and references within). In some cases, flares display circular ribbons \citep[e.g.,][]{Masson2009, Reid2012, Wang2012, Vemareddy2014, Liu2015}, or three quasi-parallel ribbons \citep{Wang2014}, or X-shaped ribbons \citep{Li2016, Liu2016b}, signifying a complicated magnetic topology.

Theoretical/numerical models have demonstrated that magnetic reconnection is intrinsically related to two classes of topological structures, separatrices \citep{Gorbachev1988, Mandrini1991} and quasi-separatrix layers (QSLs; \citealt{Demoulin1996, Demoulin1997}). Separatrices are made up of magnetic field lines passing through either a coronal null point (NP; \citealt{Lau1993}) or a bald patch (BP; \citealt{Titov1993}). Across separatrices the connectivity of field lines changes discontinuously (see \citealt{Longcope2005} for a review). Reconnection of separatrix field lines through NPs is in a pairwise fashion, which is the essential idea underlying the magnetic breakout model \citep{Antiochos1999} for coronal mass ejections (CMEs). In contrast, QSLs, quantified by the squashing factor Q~$\gg$1 \citep{Titov2002}, are thin volumes where the connectivity of field lines changes drastically but still continuously \citep{Priest1995, Demoulin1997}. Due to the strong distortion of magnetic mapping at QSLs, the generic buildup of intense electric current sheets along QSLs is expected analytically \citep{Demoulin1996a}, and confirmed later on by numerical simulations \citep[e.g.,][]{Aulanier2005, Effenberger2011, Craig2014}. In the case of QSLs, the continuous exchange of connectivity between neighboring field lines induces an apparent ``slipping" motion of field lines, which is termed ``slipping reconnection" if the speed is sub-Alfv\'{e}nic, or ``slip-running reconnection" if the speed is super-Alfv\'{e}nic \citep{Aulanier2006}. Slipping reconnection along QSLs is theoretically predicted in solar eruptions \citep{Aulanier2012, Masson2012, Janvier2013}, and expected to be more general than reconnection at separatrices where Q$\rightarrow\infty$, a limiting case of QSLs.

Apart from the long-developed theoretical concept of the slipping-type reconnection, observations of such reconnections especially those related to solar flares have been rare \citep{Aulanier2007, Testa2013, Dudik2014, Li2014, Li2015a, Dudik2016, Gou2016, Sobotka2016, Zheng2016}. This is mainly in that the high-quality observations required for an accurate depiction of slipping reconnection were only available a few years ago with ever-increasing spatiotemporal resolutions of solar instruments such as the Atmospheric Image Assembly (AIA; \citealt{Lemen2012}) on the Solar Dynamic Observatory (SDO). Compared to the propagating kernel brightenings at the footpoints of coronal loops \citep[e.g.,][]{Dudik2014, Sobotka2016}, the apparent slippage of loops appears more perceptible. However, the newly reconnected hot coronal loops may not be detectable immediately in the narrow-band EUV images in the initial energy release phase, but becoming visible later on as plasma cools down through radiative losses and thermal conduction. In this paper, ``slippage of loops" is used to describe an apparent propagation of the most noticeably heated loops in EUV passbands.

Here we present the observation of an intriguing M6.5 flare. The precursor and fine-scale structure of this flare have been discussed by \citet{Wang2017} and \citet{Jing2016}, respectively. This study focuses on the large-scale dynamics associated with the flare. Our results are in favor of a slipping-type reconnection followed by thermodynamic evolution of coronal loops.

\section{DATA AND ANALYSIS TOOLS}
We used SDO/AIA UV 1600~\AA\ images to study the flare morphology. Differential Emission Measure (DEM) analysis was performed with the six AIA EUV passbands (94, 131, 171, 193, 211, 335 \AA) using the sparse inversion method \citep{Cheung2015}. This method provides a number of benefits: (1) it has been validated against a diverse range of thermal models, (2) it returns positive definite DEM solutions, and (3) it is fast and can be used to generate sequences of DEM images. Full-disk images from \verb#aia.lev1# series at the AIA-HMI Joint Science Operations Center (JSOC) were de-rotated to a reference time set to be 18:00 UT, and co-aligned and sampled onto a common plate-scale with a pixel size of 0.6\arcsec. For our region-of-interest (ROI), DEM solutions were obtained on a pixel-by-pixel basis on a temperature grid with $\log T = 5.7, 5.8, \ldots, 7.6, 7.7$.

To study the magnetic configuration of the active region (AR), we extrapolated the 3D nonlinear force-free field (NLFFF) and potential field with the weighted optimization method \citep{Wiegelmann2004} and Green’s function method, respectively. We used \verb#hmi.sharp_cea_720s# series obtained by the SDO/Heliospheric and Magnetic Imager (HMI; \citealt{Schou2012}) as the boundary conditions. The magnetograms were re-mapped using a cylindrical equal area (CEA) projection, and presented as ($B_r$,$B_\theta$,$B_\phi$) in heliocentric spherical coordinates corresponding to ($B_z$,−$B_y$,$B_x$) in heliographic coordinates \citep{Sun2013a}. The extrapolation was performed within a box of 840$\times$448$\times$448 uniform grid points, corresponding to $\sim$300$\times$160$\times$160~Mm. We then calculated the squashing factor Q in the box volume of the potential field with the code introduced by \citet{Liu2016a}. The reason to use a potential field model instead of an NLFFF is that structural skeletons of magnetic field are very robust as demonstrated by many earlier studies \citep[e.g.,][]{Demoulin2006, Demoulin2007a, Liu2014a}.

\section{OBSERVATIONS AND ANALYSIS}
The M6.5 flare appeared in NOAA AR 12371, and was associated with a halo CME. As shown in Fig.1a, the flare emission shows three peaks (at 17:52, 17:58, and 18:12 UT) in Fermi hard X-ray (HXR) flux, and reaches maxima in GOES soft X-ray (SXR) flux at 18:23 UT. During the early and impulsive phases, the flare exhibits two ribbons in the chromosphere (denoted by R1 and R2 in Fig.1b) with the primary magnetic PIL (green curve) in between. Two HXR footpoint sources (red contours) are seen lying within R1 and R2. Then R2 displays a fast elongation motion in the southward direction. The elongated brightenings quickly evolve into a V-shaped form (Fig.1b\&1f), from which a jet stream of material erupts out abruptly near 18:12 UT. Immediately after the eruption, a $\nabla$-shaped dimming region appears in the EUV passbands, then shrinks in size and is finally closed, leaving only a dim slit open in the west (Fig.1g). The dim slit is about 60~Mm long and 6-7~Mm wide, oriented in a direction nearly perpendicular to the primary PIL between R1 and R2, and channeling the flaring region to a neighboring sunspot. Hereafter we use the term ``dimming channel" to describe this structure. The third ribbon of interest (denoted by R3 in Fig.1c\&1d) appears at approximately 18:40 UT in the eastern end of the dimming channel, then propagates along the dimming channel to its western end (Fig.1c\&1d). During this period, R2 gradually fades away. Two HXR footpoint sources which were initially associated with R1/R2 ribbon pair are now transitioned to R1/R3 pair (Fig.1d). The propagation of R3 is followed by the apparent slippage of loops, particularly notable at their western ends in the negative polarities (blue circles in Fig.1h\&1i).

Four slits, denoted by S1$-$S4 in Fig.2a, are set to construct the distance-time stackplots displayed in Fig.2b$-$2d. The section in Fig.2b constructed using S1 shows the evolution of the $\nabla$-shaped dimming region. Fig.2c is a composite stackplot, in which we used S2 to detect the motion of the jet-like eruption, and S3 to detect the propagation of R3. When the $\nabla$-shaped dimming region reaches its maximum areal extent near 18:12 UT, the jet-like ejection takes place abruptly and travels at a projected speed of 217~km s$^{-1}$ from the dimming region to a remote site southeastward to the AR. At about 18:40 UT, R3 appears and propagates westward. The speed of R3 propagation is not uniform $--$ it is relatively slow (4-6~km s$^{-1}$) at the beginning and at the end of the propagation in stronger magnetic field ($B_z\sim$ -580~G), and faster (23~km s$^{-1}$) in between when crossing over weaker field ($B_z\sim$ -140~G). In Fig.2d, the distance-time stackplot constructed with S4 shows the apparent slippage of the western legs of coronal loops. The apparent slippage is actually along the path of R3, but the position of S4 is slightly displaced from that of S3 to avoid the influence of R3 on the detection of loops. Through S4 the apparent slippage of loops starts around 20:00 UT, $\sim$80~min after the onset of R3 propagation and with a rather uniform speed (4~km s$^{-1}$).

Fig.3 shows magnetic field and topology of the pre-flare AR. In the NLFFF (Fig.3b), an inner flux-rope-like structure resides between R1 and R2, the rise of which and the subsequent reconnection are presumably pertinent to the observed ribbon pair R1/R2 on the onset of the flare. Note that the field lines in the potential field (Fig.3c) show a well organized pattern $--$ those emanating from the southern portion of R1 are more compact and connecting to R2, whereas those emanating from the northern portion are progressively higher and gradually landing westward. The diverging locations of the footpoints are more readily matched to R3. This potential magnetic configuration schematically explains the transition from R1/R2 to R1/R3 as the reconnection proceeds northward and toward higher altitudes.

Fig. 4a shows the photospheric slogQ map, where $slogQ = sign (Bz) log_{10}Q$ \citep{Titov2011}. The high-Q structures outline the footprints of the prominent QSLs. Although the slogQ map here is rather complicated with many small-scale features, we see two large-scale high-Q structures in the AR complex: a C-shaped one on the lefthand side of the map in the positive polarities (denoted by C), and a roughly transverse Y-shaped one in the negative polarities (denoted by transverse Y). The two arms of the Y interlock loosely with the C, and the trunk of the Y is composed of two high-Q lines extending to the right. To compare with the ribbon morphology, this slogQ map was blended with the maps of AIA 1600~\AA\ (Fig. 4c) and 211~\AA\ (Fig. 4d). All the maps were acquired at approximately the same time, 19:36 UT. At this time, R2 is already fading away, whereas R1 and R3 are still seen nearly on their full extent. We see that R1 is roughly matched by a segment of the C, while R3 coincides well with the trunk of the Y, especially the top high-Q line of the Y. We traced the potential field lines from the top high-Q line of the Y. The conjugate footpoints of these field lines land at the same place of the C where R1 is situated (Fig. 4e). Such a connectivity of the field lines bears a similarity to the apparent slipping loops seen in the AIA EUV passbands.

Fig. 5 shows the DEM maps at 19:00 UT. A time series of such DEM maps is available in the online journal. In the Movie, one can clearly identify the apparent propagation of a plasma bulk over the period of 18:40$-$19:40 UT, which is co-spatial and co-temporal with the R3 propagation, and accompanied by a HXR footpoint source (Fig.5h$\&$5i). The propagation of the plasma bulk is best seen in log T/K $\in$ [5.95,6.25], and also broadly visible at other temperature bins up to [6.85,7.15]. The thermodynamic evolution of loops is also demonstrated in Movie~3. The composite DEM image of log T/K $\in$ [6.55,6.85] (green) and [6.85,7.15] (red) shows a clear color transition from cooler loops to hotter ones, in a direction consistent with the apparent slippage of loops.

\section{DISCUSSION AND CONCLUSION}

The flare starts as a typical two-ribbon flare (R1/R2), followed by three episodes related to the evolution of R3: a jet-like ejection, the formation of a dimming channel, and the propagation of R3 along the dimming channel. The potential field model as shown in Fig. 3 schematically illustrates the spatial relationship among the flare loops. It appears that R1 and R2 are connected by the low-lying field lines above the inner flux rope, whereas R1/R3 connection involves higher field lines, the connectivity of which are continuous while strongly diverging. Furthermore, it is found that: (1) the path of R3 propagation shows a good spatial correlation with a portion of prominent QSL footprints, and (2) the observed R1/R3 loop system can be reproduced by the linkage of a set of modeled field lines related to the QSLs. The results are strongly in favor of a slipping-type reconnection for R3.

Presumably the context scenarios are as follows. When the inner flux rope starts to rise, magnetic reconnection occurs and produces a two-ribbon flare R1/R2. As the reconnection proceeds toward higher altitudes and cross the QSLs, slipping reconnection sets in and contributes to the ribbon pair R1/R3. Prior to the transition to the slipping reconnection, a jet-like eruption occurs nearby. Its relation to the flare is not clear. Here we speculate that this jet may alleviate constraints, by pushing aside some of overlying field lines, to facilitate the transition. As the ejecta leave the corona, a full or partial opening of magnetic field occur that lead to depletion of plasma and a corresponding EUV intensity dimming underneath. The dimming channel evolves from the region associated with the ejection, and represents footpoints of field lines that are opened up by reconnections associated with the jet. The apparent slippage of coronal loops manifests the interplay between successive fading of old loops and the appearance of new ones during the cooling process of the reconnection, in the same manner as the apparent growth of post-flare loops commonly seen in two-ribbon flares.

In comparison with the previously reported slipping reconnection events, this one proceeds across a particularly long distance ($\sim$60 Mm) over a long period of time ($\sim$50 min). The propagation speed of R3 suggests that the reconnection is in the regime of slow slipping rather than the slip-running at super-Alfv\'{e}nic speed. The 80~min time delay between the propagation of R3 and the apparent slippage of loops $\sim$80~min is probably with respect to the loop length.

The shape of QSL footprints in this event (Fig. 4) is similar to that in the MHD simulation by \citet{Aulanier2005} and as depicted in their Fig.3 where two strong high-Q footprints whirl around each other in a quadrupolar magnetic configuration in a similar way as the C and the top high-Q line of the Y do in this case (see Fig. 4). In other cases of simulation, two high-Q footprints display an asymmetric double-J shaped pattern \citep[e.g.,][]{Aulanier2012, Janvier2013}. These simulations demonstrate that QSLs are preferable sites for the formation of intense current sheets and reconnection, much like we noted for the spatial coincidence between the QSL footprints and the ribbon pair R1/R3. In both our observation and the simulations, the speed of the slipping reconnection is not uniform. It is yet unclear whether the magnetic field plays a role in determining the reconnection speed, but there is seemingly a correlation between the two, which, of course, requires further investigation.


The DEM analysis provides important information on the thermal behavior of the flare which corroborates the general reconnection scenario. The magnetic reconnection and the resulting flare produce high-temperature plasma initially. In this case, the loops in the core region can be heated to over 10 MK. The plasma temperature of R3 is found to be widely distributed from 0.6~MK up to 10~MK. The presence of hot plasma at temperatures around 10~MK is presumably attributed to the ``chromospheric evaporation" process driven by pressure balance across the transition region \citep{Hirayama1974}. Evaporation fills the reconnected loops with $>\sim$~10 MK plasma that have been heated at the footpoints, then these loops gradually cools down.

To summarize, the observation showcases a spectacular propagation of a flare ribbon that indicates a large-scale, long-duration, slipping-type reconnection. The observation reveals two distinct phases: the propagation of footpoints driven by nonthermal particle injection, and the apparent slippage of loops governed by radiative plasma cooling. The key feature of this study is that the thermal evolution of the slipping reconnection has been explored. As a result we found the considerable differences, in timing and speed, between the footpoint propagation and the apparent loop slippage.

\acknowledgments We thank the NASA SDO team for HMI and AIA data. HMI and AIA are instruments on board SDO, a mission for NASA's Living with a Star program. J.J., Y.X., C.L. and H.W. were supported by NASA grants NNX13AG13G, NNX13AF76G, NNX14AC12G, NNX14AC87G, NNX16AL67G and NNX16AF72G, NSF grants AGS 1250374, 1348513, 1408703, and 1539791. R.L. acknowledges the support by NSFC 41474151 and the Thousand Young Talents Program of China. M.C.M.C. acknowledges support by NASA contract NNG04EA00C (SDO/AIA) and grant NNX14AI14G (Heliophysics Grand Challenges Research). J.L. thanks ISEE of Nagoya University and KOFST for their generous support.


\begin{figure}
\begin{center}
\epsscale{1.} \plotone{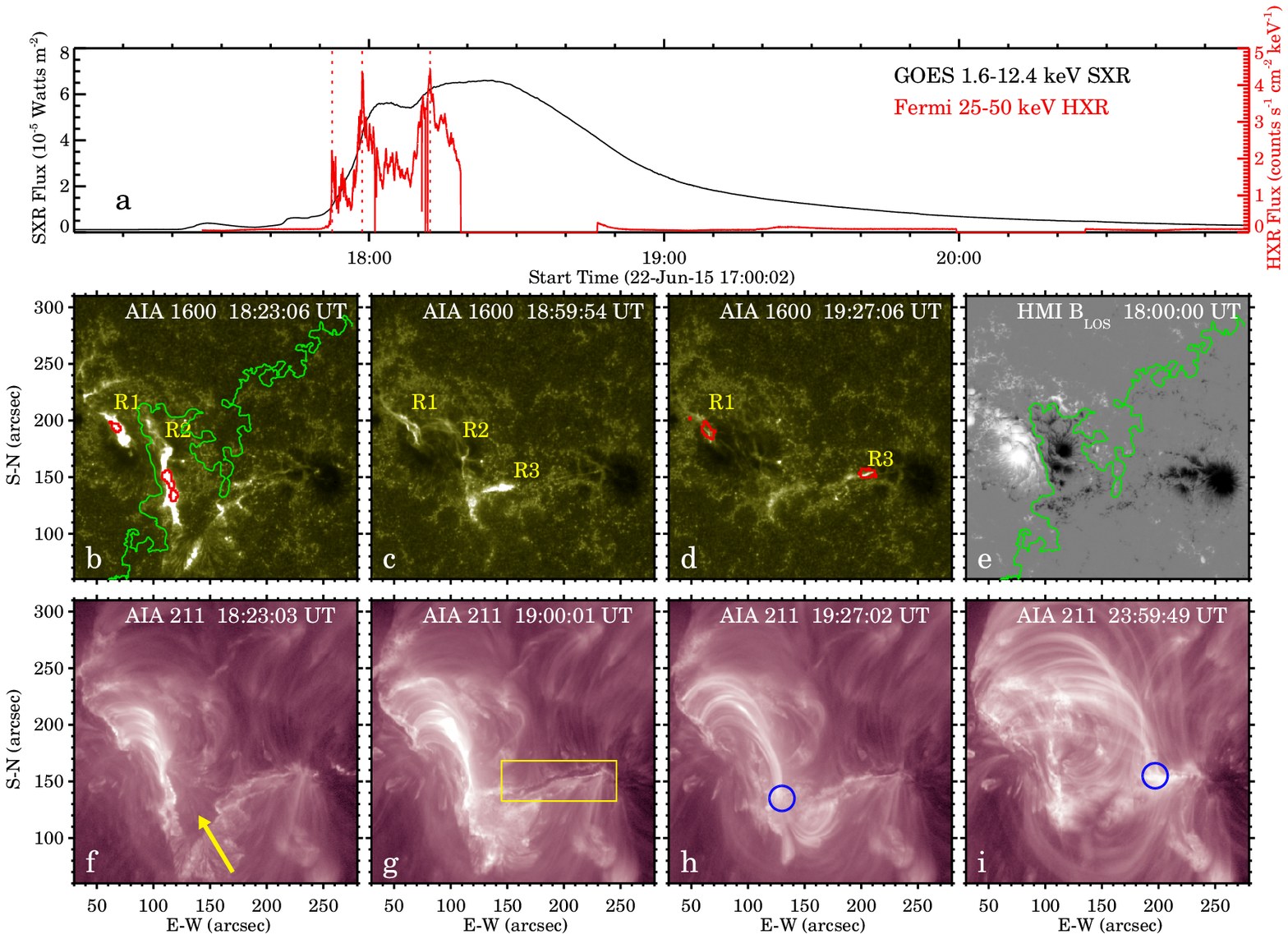} \caption{(a): Lightcurves of GOES SXR flux (black) and Fermi HXR flux (red; not available during 18:19-18:48 UT and 20:00-20:25 UT). Three vertical dashed lines mark the three Fermi HXR peaks. (b-d): AIA 1600~\AA\ images showing three flare ribbons R1, R2 and R3. The green curves delineate the magnetic PILs. The red contours show RHESSI HXR intensity integrated in the time intervals b (18:22:30 - 18:23:30 UT) and d (19:26:30 - 19:27:30 UT), in the 25-50 keV energy range. The contour level is 60\% of the intensity maximum. (e): HMI line-of-sight (LOS) magnetogram. (f-i): AIA 211~\AA\ images showing the evolution of coronal configuration. The arrow points to a $\nabla$-shaped dimming region caused by a jet (not shown in this figure). The box in (g) encloses the dimming channel. The western footpoints of the loops (blue circles) show an apparent slipping motion. An animation is available in the online journal.}
\label{fig01}
\end{center}
\end{figure}


\begin{figure}
\begin{center}
\epsscale{1.} \plotone{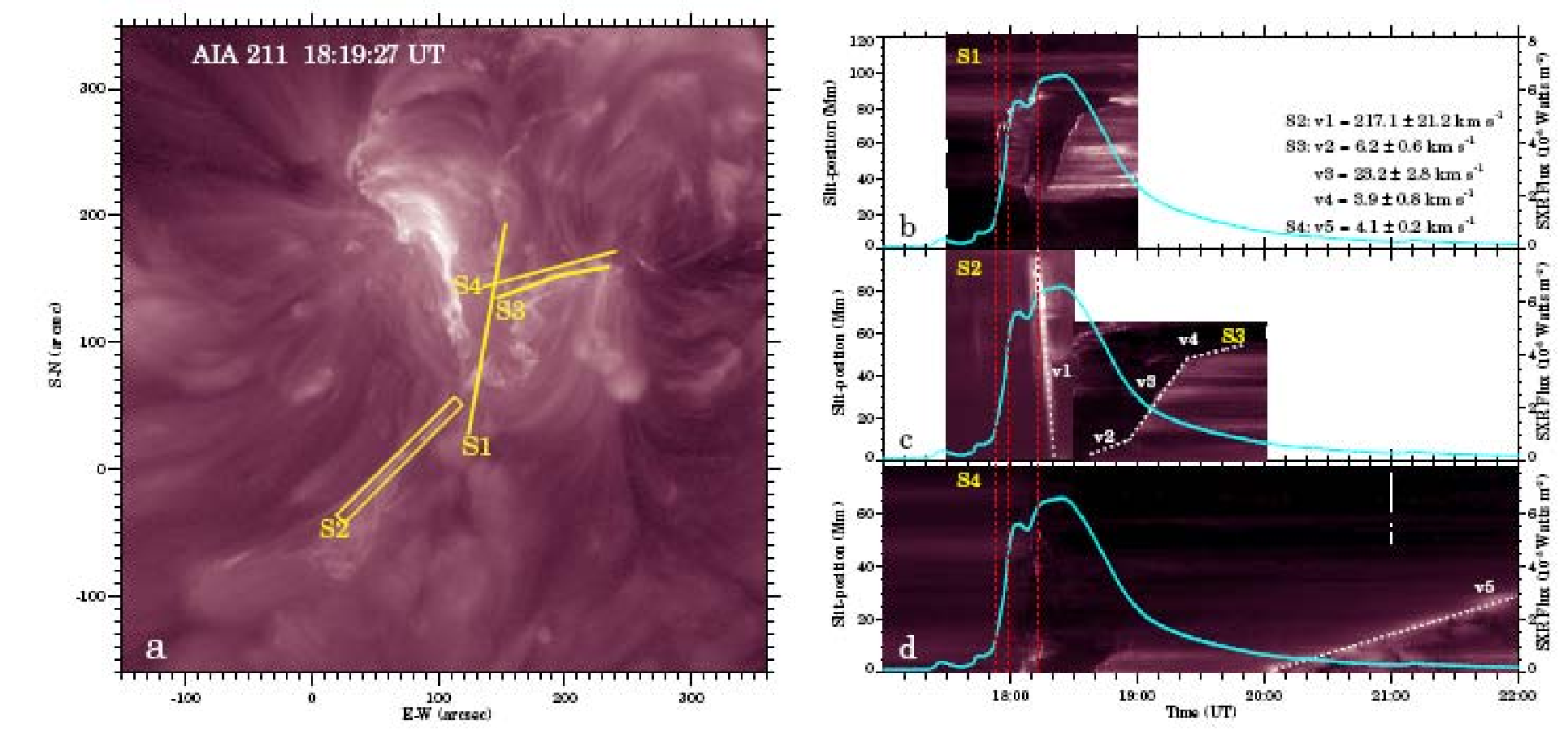}
\caption{(a): an AIA 211~\AA\ snapshot, with four slits (S1-S4) to construct the distance-time stackplots displayed in the right panels. In all cases, the end with the label (S1-S4) corresponds to the slit-position ``0". For the stripe-like S2, intensities at each time are averaged over the 15 pixels of the stripe width. (b-d): the distance-time stackplots, overlaid with the GOES SXR lightcurve. Three vertical red dashed lines mark the Fermi HXR peaks. The vertical white line near 21:00 UT in (d) was caused by a 84~s data gap. The linear fitting functions are overplotted with the white dashed lines, and the fitted speeds are given in (c). An animation is available in the online journal. } \label{fig02}
\end{center}
\end{figure}

\begin{figure}
\begin{center}
\epsscale{0.6} \plotone{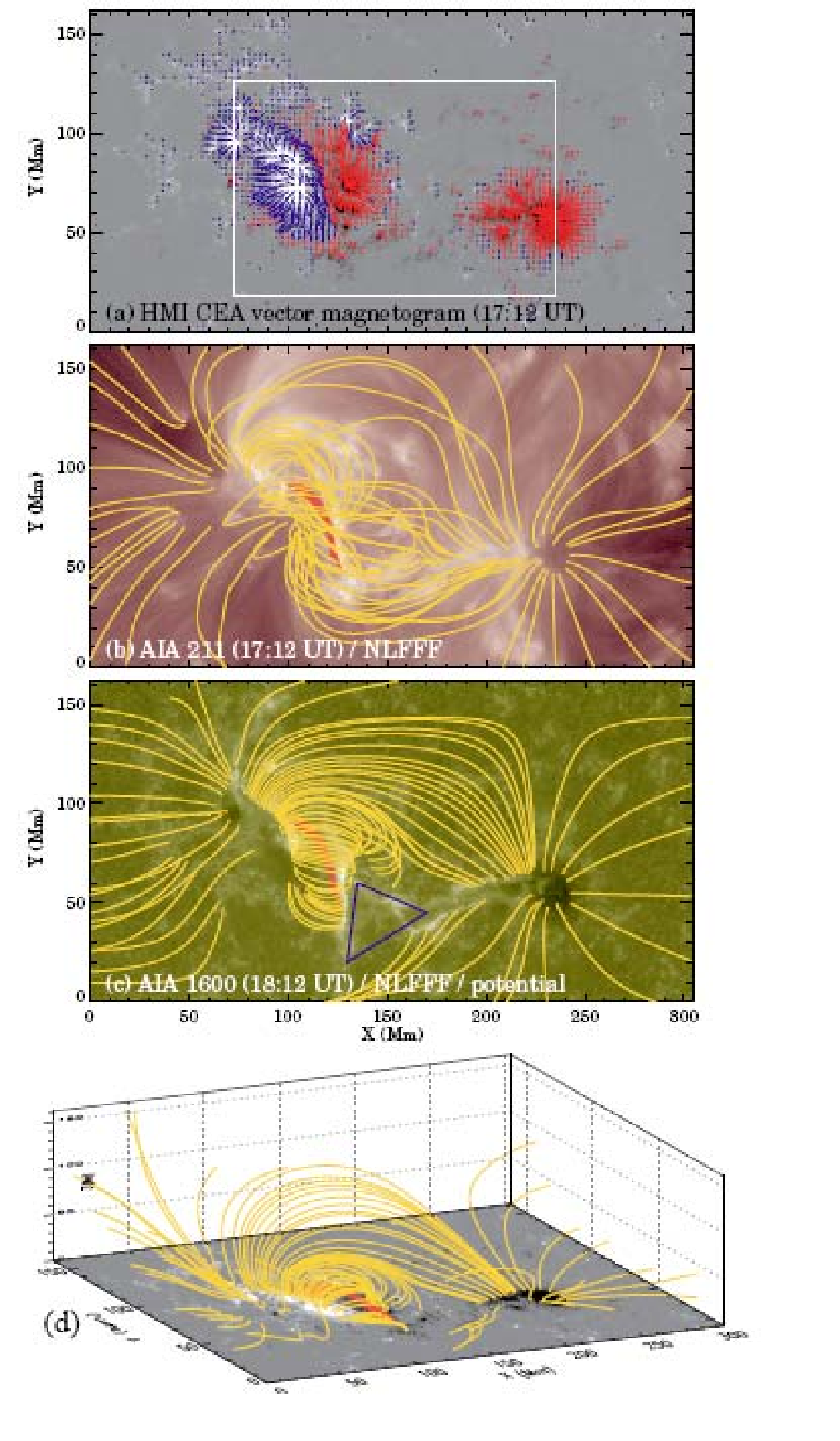}
\caption{(a) HMI vector magnetogram based on which an NLFFF and a potential field are extrapolated. On the background $B_z$ image, red/blue arrows show the horizontal components originating from negative/positive $B_z$. The rectangle indicates the field-of-view (FOV) of the slogQ map in Fig.4. (b) NLFFF lines superimposed on an AIA 211~\AA\ image. The inner flux rope is displayed in red. (c) Potential field lines (yellow) and the inner flux rope (red) shown in (b), superimposed on an AIA 1600~\AA\ image. The blue triangle indicates the location of the $\nabla$-shaped dimming region. (d) Side view of the flux rope (red) and potential field lines (yellow).}
\label{fig03}
\end{center}
\end{figure}

\begin{figure}
\begin{center}
\epsscale{1.} \plotone{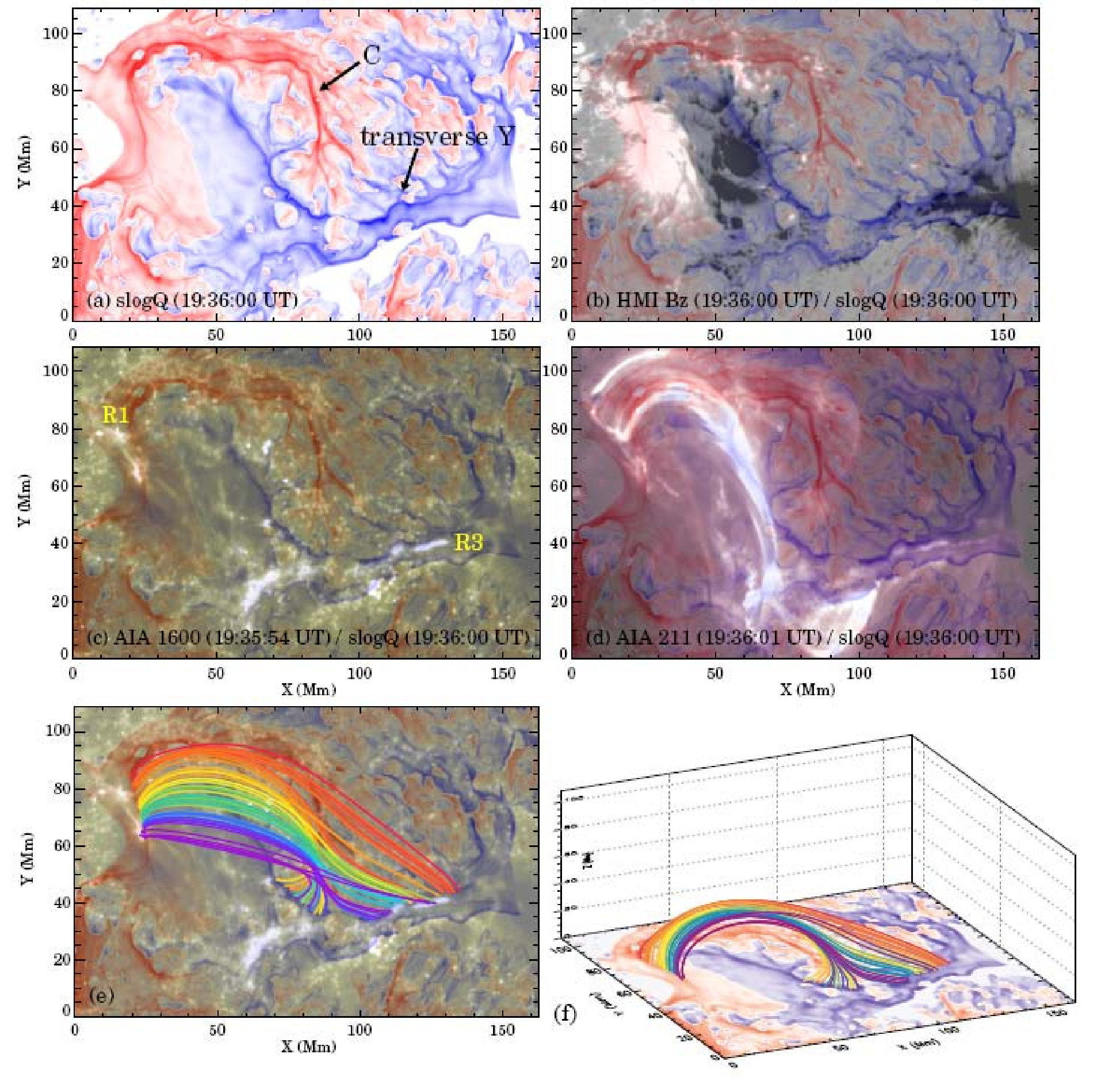}
\caption{(a): The slogQ map at z = 0 (photosphere) scaled between $-/+$5 (blue/red). The white regions are related to field lines which are open when reaching the lateral/top boundaries of the computational domain. (b-d) The slogQ map blended with the maps of HMI $B_z$, AIA 1600~\AA\ and 211~\AA. All maps are remapped with the CEA projection. (e-f): Top and side views of potential magnetic field lines (colored curves) which are traced starting from the top high-Q line of the Y.} \label{fig04}
\end{center}
\end{figure}

\begin{figure}
\begin{center}
\begin{tabular}{ll}
\resizebox{160mm}{!}{\includegraphics{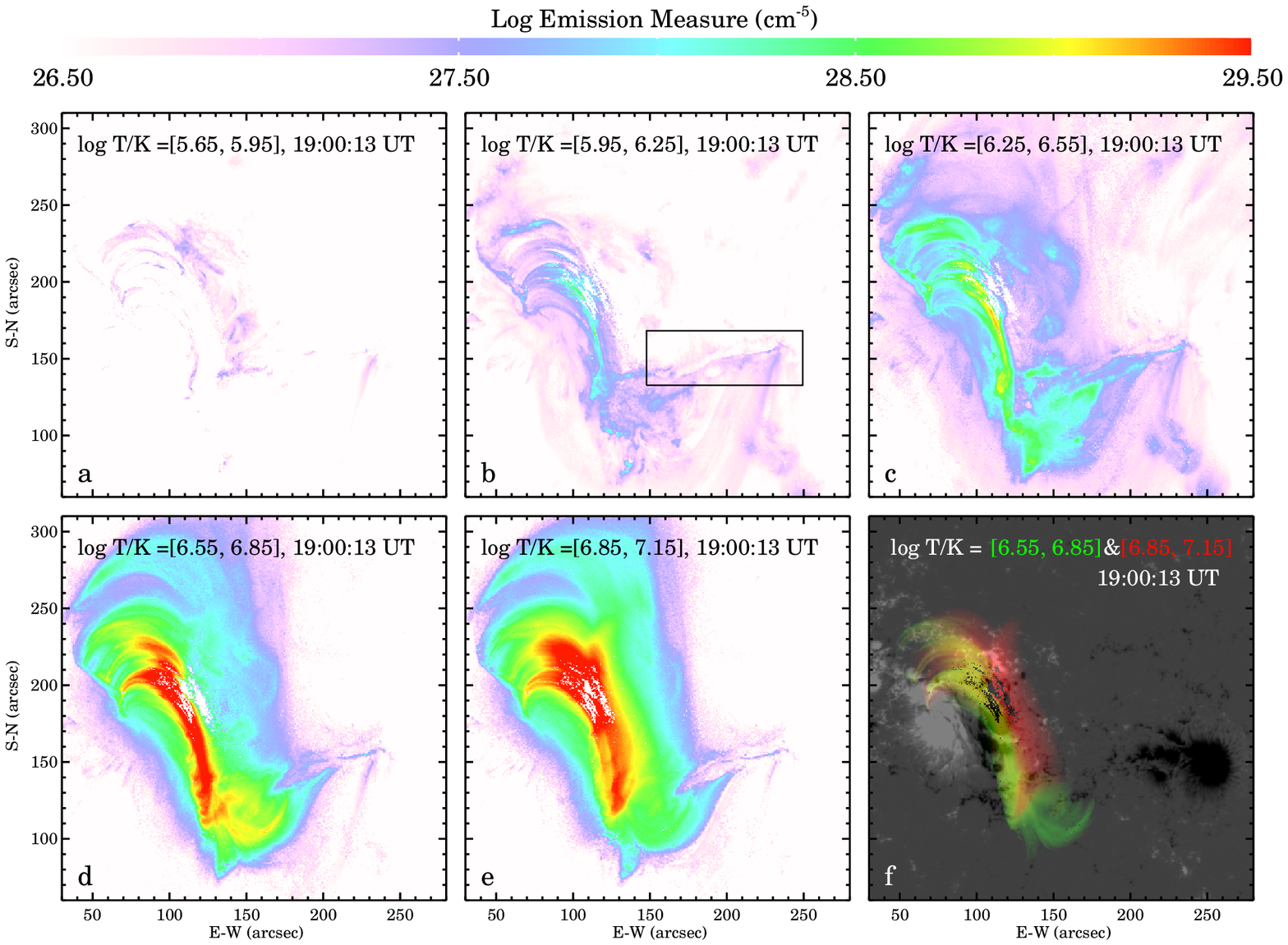}}\\
\resizebox{160mm}{!}{\includegraphics{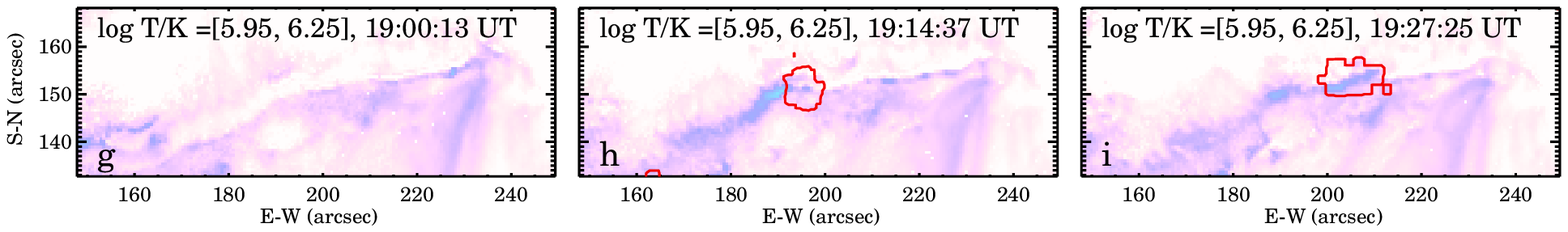}}\\
\end{tabular}
\caption{(a-e): DEM maps of $\sim$19:00 UT. The color-coding indicates the total emission measure contained within a log T range indicated in each panel. The box in (b) shows the FOV for (g-i). (f): A composite image of a HMI LOS magnetogram (background) with the DEM maps of log T/K $\in$ [6.85,7.15] (red) and [6.55,6.85] (green), all acquired at $\sim$19:00 UT. (g-i): The EM maps of 19:00, 19:14 and 19:27 UT, in the temperature range log T/K $\in$ [5.95,6.25]. The red contours show 60\% of the RHESSI HXR maximum intensity, integrated in the 25-50 keV energy range over the time intervals h (19:13:30 - 19:16:30 UT) and i (19:26:30 - 19:27:30 UT). An animation is available in the online journal. } \label{fig05}
\end{center}
\end{figure}

\end{document}